\documentclass[12pt,reqno]{amsart}

\usepackage{amsmath,
amsthm,
cite
}

\usepackage{a4wide}

\usepackage{times}

\usepackage{psfrag}
\usepackage[dvips]{graphicx}
\usepackage{epsf,epsfig}
\numberwithin{equation}{section}
\newtheorem{theorem}{Theorem}
\newtheorem{prop}[theorem]{Proposition}
\newtheorem{coro}[theorem]{Corollary}
\newtheorem{lemma}[theorem]{Lemma}

\DeclareMathOperator{\sh}{sh}
\DeclareMathOperator{\ch}{ch}

\usepackage{titlesec}
\titleformat{\section}
{\bfseries\scshape\centering}
{\thesection.}{.5em}{}
\titleformat{\subsection}
{\rmfamily\bfseries}
{\thesubsection}{.5em}{}

\begin{document}
\baselineskip 19pt
\parskip 7pt
\sloppy


\title[$L$-function,  Slater's Identities, and Quantum Invariant]{
  Hypergeometric Generating Function of $\boldsymbol{L}$-function,
  Slater's
  Identities, and
  Quantum   Invariant 
}


\author[K. Hikami]{Kazuhiro \textsc{Hikami}}

\address{Department of Physics, Graduate School of Science,
  University of Tokyo,
  Hongo 7--3--1, Bunkyo,
  Tokyo 113--0033, Japan.
}

    \urladdr{http://gogh.phys.s.u-tokyo.ac.jp/{\textasciitilde}hikami/}

\email{\texttt{hikami@phys.s.u-tokyo.ac.jp}}

 \author[A. N. Kirillov]{Anatol N. Kirillov}

\address{
Research Institute  of Mathematical Sciences
(RIMS),
 Kyoto University, 
  Kyoto 606-8502, Japan
}

\urladdr{http://www.kurims.kyoto-u.ac.jp/{\textasciitilde}kirillov/}
\email{\texttt{kirillov@kurims.kyoto-u.ac.jp}}


\dedicatory{
  Dedicated to Ludwig Dmitrievich Faddeev on the occasion of his
  seventieth     birthday
}


\date{\today}


\begin{abstract}

We study certain connections between the quantum invariants of the torus knots
$\mathcal{T}_{3,2^k}$ and some $q$-series identities.
In particular, we obtain new generalizations of
Slater's identities~(83) and~(86).

\end{abstract}



\subjclass[2000]{
11B65,
57M27,
05A30,
11F23
}


\maketitle
\section{Introduction}

Recent studies reveal an intimate connection between the quantum
invariants for 3-manifold and the modular forms.
This remarkable observation originates  from
Refs.~\citen{LawreRozan99a,LawrZagi99a}.
In a slightly different context~\cite{DZagie01a},
it was discussed that the generating function of
Stoimenow's  upper bound of the number of the Vassiliev
invariant coincides with the half-differential of the Dedekind
$\eta$-function.
{}From a viewpoint of the quantum invariant, this generating function
happens to coincide with Kashaev's invariant~\cite{Kasha95}
for trefoil.
This quantum knot invariant
was originally defined by use of the quantum dilogarithm
function~\cite{FaddKash94},
and is
a specific value of the
colored Jones polynomial~\cite{MuraMura99a}.
Motivated by this coincidence, it was
clarified~\cite{KHikami02b,KHikami02c,KHikami03a,KHikami03c} that
Kashaev's
invariant for torus knot $\mathcal{T}_{s,t}$ and torus link
$\mathcal{T}_{2,2 m}$
respectively
coincides with the Eichler integral of the Virasoro character for the
minimal model $\mathcal{M}(s,t)$ and that of the
$\widehat{su}(2)_{m-2}$ character.

One of the  benefits of the correspondence between the character and the
quantum invariant is that we can find new $q$-series identities.
For example, it is well known that
the Virasoro character for $\mathcal{M}(2,2\, m+1)$
is related to the
Gordon--Andrews identity (generalization of the famous
Rogers--Ramanujan identity).
Motivated from an explicit form of
Kashaev's invariant for torus knot $\mathcal{T}_{2,2m+1}$,
we obtained a new $q$-series which may be regarded as a one-parameter extension of the Gordon--Andrews identity~\cite{KHikami02c}.
A case of $m=1$ corresponds to Zagier's $q$-series identity~\cite{DZagie01a}.
See also  Refs.~\citen{AndrUrroOnok01a,LoveKOno02a,CoogKOno02a,KOno04Book}
for other generalizations of Zagier's identity.
Therein constructed were
the $q$-hypergeometric type generating functions of the
$L$-function at the negative integers.
Purpose of this article is to   propose a $q$-series identity which is
related to  Slater's  identities.


We define the Santos polynomial~\cite{AndreSanto97a,AndKnoPauPro01a} by
\begin{subequations}
  \begin{align}
    S_n(q) & =
    \sum_{k=0}^{\lfloor n/2 \rfloor} q^{2 k^2} \,
    \begin{bmatrix}
      n \\
      2 \, k
    \end{bmatrix} ,
    \\[2mm]
    T_n(q) & =
    \sum_{k=0}^{\lfloor (n-1)/2 \rfloor} q^{2 k (k+1)} \, 
    \begin{bmatrix}
      n \\
      2 \, k +1
    \end{bmatrix} .
  \end{align}
\end{subequations}
Definitions of our notation  are  summarized in  Section~\ref{sec:notation}.
These polynomials can also be defined recursively by
\begin{equation}
  \begin{pmatrix}
    S_{n+1}(q) \\[2mm]
    T_{n+1}(q)
  \end{pmatrix}
  =
  \begin{pmatrix}
    1 & q^{n+1} \\[2mm]
    q^n & 1
  \end{pmatrix}
  \,
  \begin{pmatrix}
    S_n(q) \\[2mm]
    T_n(q)
  \end{pmatrix} ,
\end{equation}
with an initial condition
\begin{equation*}
  \begin{pmatrix}
    S_0(q) \\[2mm]
    T_0(q)
  \end{pmatrix}
  =
  \begin{pmatrix}
    1 \\[2mm]
    0
  \end{pmatrix}  .
\end{equation*}
By use of  the Santos polynomials,
we define the formal $q$-series $X^{(0)}(q)$  and  $X^{(1)}(q)$ by
\begin{subequations}
  \label{define_X}
  \begin{align}
    X^{(0)}(q)
    & =
    \sum_{n=0}^\infty (q)_n \, 
    \left( T_n(q) + T_{n+1}(q) \right) ,
    \\[2mm]
    X^{(1)}(q)
    & =
    \sum_{n=0}^\infty  (q)_n \,
    \left( S_n(q) + S_{n+1}(q) \right) .
  \end{align}
\end{subequations}
To state one of our main theorems, we introduce the periodic functions as
follows;
\begin{subequations}
  \label{define_chi}
  \begin{gather}
    \begin{array}{c|ccccc}
      n \mod 24 & 5 & 11 & 13 & 19  & \text{others}
      \\
      \hline      \hline
      \chi_{24}^{(0)}(n) & 1 & -1 & -1 & 1 & 0
    \end{array}
    \\[2mm]
    \begin{array}{c|ccccc}
      n \mod 24 & 1 & 7 & 17 & 23  & \text{others}
      \\
      \hline       \hline
      \chi_{24}^{(1)}(n) & 1 & -1 & -1 & 1 & 0
    \end{array}
  \end{gather}
\end{subequations}

\begin{theorem}
  \label{theorem:Slater}
  Let $X^{(a)}(q)$ for $a=0,1$ be defined by eqs.~\eqref{define_X}.
  We have an asymptotic expansion in $z \searrow 0$ as
  \begin{subequations}
    \begin{align}
      X^{(0)}(\mathrm{e}^{-z})
      & =
      \mathrm{e}^{25 \, z /48} \,
      \sum_{n=0}^\infty
      \frac{t_n^{(0)}}{n!} \,
      \left( \frac{z}{48} \right)^n ,
      \\[2mm]
      X^{(1)}(\mathrm{e}^{-z})
      & =
      \mathrm{e}^{ z /48} \,
      \sum_{n=0}^\infty
      \frac{t_n^{(1)}}{n!} \,
      \left( \frac{z}{48} \right)^n .
    \end{align}
  \end{subequations}
  Here $t$-series is given in terms of the $L$-function associated
  with $\chi_{24}^{(a)}(n)$;
  \begin{align}
    t_n^{(a)}
    & =
    \frac{1}{2} \, (-1)^{n+1} \, L(-2 \, n -1 , \chi_{24}^{(a)} )
    \nonumber
    \\
    & =
    \frac{1}{2} \, (-1)^n \, 
    \frac{24^{2 n+1}}{2 \, n+2} \,
    \sum_{k=1}^{24} \chi_{24}^{(a)}(k) \, B_{2 n+2}(k/24) ,
    \label{define_t_12}
  \end{align}
  where $B_n(x)$ is the $n$-th Bernoulli polynomial.
\end{theorem}

We note that  generating functions of the $t$-series  are written as
\begin{align*}
  \frac{\sh(3 \, x) \, \sh(4 \, x)}{\sh(12 \, x)}
  & =
  \frac{1}{2}
  \sum_{n=0}^\infty
  \chi_{24}^{(0)}(n) \, \mathrm{e}^{- n \, x}
  =
  \sum_{n=0}^\infty (-1)^n \frac{t_n^{(0)}}{(2 \, n+1)!} \, x^{2 n +1} ,
  \\[2mm]
  \frac{\sh(3 \, x) \, \sh(8 \, x)}{\sh(12 \, x)}
  & =
  \frac{1}{2}
  \sum_{n=0}^\infty
  \chi_{24}^{(1)}(n) \, \mathrm{e}^{- n \, x}
  =
  \sum_{n=0}^\infty (-1)^n \frac{t_n^{(1)}}{(2 \, n+1)!} \, x^{2 n +1} .
\end{align*}
Some of the $t$-series are explicitly given as follows;
\begin{equation*}
  \begin{array}{c|rrrrrrr}
    n & 0 & 1 & 2 & 3 & 4 & 5 & 6
    \\
    \hline
    t_n^{(0)}
    & 1 & 119 & 37201 & 23181479 & 24453497761 & 39286795847639 &
    89443016674567921
    \\
    t_n^{(1)}
    & 2 & 142 &  38882 & 23439022 & 24521135042 & 39313934084302 &
    89458458867741602
    \\
    t_n^{(2)}
    &  2  & 184 & 53792 & 32965504 & 34630287872 & 55579108685824 &
    126502446478794752
  \end{array}
\end{equation*}
See eq.~\eqref{define_t_etc} for  definition of the  $t$-series
$t_n^{(2)}$.

This paper is organized as follows.
In Section~\ref{sec:notation} we collect notations and identities of
$q$-series used in this paper.
See \emph{e.g.} Ref.~\citen{Andre76}.
In Section~\ref{sec:proof} we prove Theorem~\ref{theorem:Slater}.
In Section~\ref{sec:Knot} we study a (nearly)
modular property of our
$q$-series.
We show that $X^{(a)}(q)$ is regarded as the Eichler integral of the
modular form with weight $1/2$, which corresponds to the character of the Virasoro
minimal model $\mathcal{M}(3,4)$.
We further discuss on a relationship with the quantum knot invariant
for the torus knot $\mathcal{T}_{3,4}$.
In Section~\ref{sec:application} we show several $q$-hypergeometric
type expression of the character of the Virasoro minimal model
$\mathcal{M}(3,k)$.

\section{Notation and Identities}
\label{sec:notation}

For our later convention,
we give a list of notations and  useful identities
(see, \emph{e.g.}, Ref.~\citen{Andre76});
\begin{itemize}

\item $q$-product and $q$-binomial coefficient (the Gaussian
  polynomial),
  \begin{gather*}
    (a)_n = (a; q)_n = \prod_{k=1}^n (1 -a \, q^{k-1}) ,
    \\[2mm]
    (a_1, a_2, \dots, a_k ; q)_n = (a_1 ;q)_n \, (a_2; q)_n \cdots
    (a_k; q)_n,
    \\[2mm]
    \begin{bmatrix}
      n \\
      k
    \end{bmatrix}
    =
    \begin{bmatrix}
      n \\
      k
    \end{bmatrix}_q
    =
    \begin{cases}
      \displaystyle
      \frac{(q)_n}{(q)_k \, (q)_{n-k}} ,
      &
      \text{for $n \geq k \geq 0$,}
      \\[3mm]
      0 ,
      &
      \text{otherwise} ,
    \end{cases}
  \end{gather*}
\item  $q$-binomial formula,
  \begin{equation}
    \label{q_binomial}
    \sum_{n=0}^\infty \frac{(a)_n}{(q)_n} \, z^n
    =
    \frac{ (a \, z)_\infty }{(z)_\infty} ,
  \end{equation}

\item 
  $q$-expansion,
  \begin{gather}
    (z)_n = \sum_{k=0}^n 
    \begin{bmatrix}
      n \\
      k
    \end{bmatrix}
    \,
    (-z)^k \, q^{k(k-1)/2} ,
    \label{q_product_and_Gauss}
    \\[2mm]
    \frac{1}{(z)_n}
    =
    \sum_{k=0}^\infty
    \begin{bmatrix}
      n+k-1 \\
      k
    \end{bmatrix}
    \,
    z^k ,
  \end{gather}

\item the Euler identity (a limit $n\to \infty$ of
  eq.~\eqref{q_product_and_Gauss}),
  \begin{equation}
    \sum_{n=0}^\infty \frac{q^{n (n-1)/2}}{(q)_n} \, z^n
    =
    ( -z)_\infty ,
  \end{equation}

\item
  the Jacobi triple product identity,
  \begin{equation}
    \label{Jacobi}
    \sum_{k\in \mathbb{Z}}
    (-1)^k \, q^{ k^2 / 2} \,  x^k
    =
    (q , 
    x^{-1} q^{\frac{1}{2}}  ,
    x \,  q^{\frac{1}{2}} ;  q)_\infty  ,
  \end{equation}

\item
  the Watson quintuple identity,
  \begin{equation}
    \label{quintuple}
    \sum_{k \in \mathbb{Z}}
    q^{ k (3 k-1) / 2} \, x^{3 k} \,
    ( 1- x \, q^k)
    =
    (q, x, q \, x^{-1} ; q)_\infty \,
    (q \, x^2, q \, x^{-2} ; q^2)_\infty
  \end{equation}

\end{itemize}
We note that
the Watson identity can be proved by use of eq.~\eqref{Jacobi}
(see \emph{e.g.},
Ref~\citen{Baile51a}).

Hereafter we use properties of the $L$-function associated with the
periodic function.
\begin{lemma}
  \label{lemma:L_function}
  Let $C_f(n)$ be a periodic function with mean value $0$ and modulus $f$.
  Then     asymptotic expansion  in a limit  $t \searrow 0$ is as
  follows;
  \begin{gather}
%
    \sum_{n=0}^\infty
    n \, C_f(n) \,  \mathrm{e}^{- n^2 t}
    \simeq
    \sum_{k=0}^\infty L(-2 \, k -1 , C_f) \, \frac{(-t)^k}{k!} ,
  \end{gather}
  where
  $L(k,C_f)$ is the $L$-function associated with $C_f(n)$,
  and is given by
  \begin{equation*}
    L(-k, C_f)
    =
    - \frac{f^k}{k+1} \,
    \sum_{n=1}^f
    C_f(n) \, B_{k+1} \left( \frac{\ n \ }{f} \right) .
  \end{equation*}
\end{lemma}

\begin{proof}
  It is a standard result using the Mellin transformation.
  See, \emph{e.g.}, Ref.~\citen{LawrZagi99a}.
\end{proof}

\section{\mathversion{bold}
  Proof of Theorem~\ref{theorem:Slater}}
\label{sec:proof}

We define functions $H^{(a)}(x) \equiv H^{(a)}(x ; q)$ by
\begin{subequations}
  \label{define_H}
  \begin{align}
    H^{(0)}(x) & =
    \sum_{n=0}^\infty \chi_{24}^{(0)}(n) \, q^{\frac{n^2 - 25}{48}} \,
    x^{\frac{n-5}{2}}
    \\
    & =
    1 - q^2 \, x^3 - q^3 \, x^4 + q^7 \, x^7 + q^{17} \, x^{12} - q^{25}
    \, x^{15} - q^{28} \, x^{16} + \cdots  ,
    \nonumber
    \\[2mm]
    H^{(1)}(x) & =
    \sum_{n=0}^\infty \chi_{24}^{(1)}(n) \, q^{\frac{n^2 - 1}{48}} \,
    x^{\frac{n-1}{2}}
    \\
    & =
    1 - q \, x^3 - q^6 \, x^8 + q^{11} \, x^{11} + q^{13} \, x^{12} - q^{20}
    \, x^{15} - q^{35} \, x^{20} + \cdots  ,
    \nonumber
  \end{align}
\end{subequations}
where the periodic function $\chi_{24}^{(a)}(n)$ is defined in eq.~\eqref{define_chi}.
We easily see
that these $q$-series solve the
following $q$-difference equations;
\begin{equation}
  \begin{aligned}
    H^{(0)}(x)
    & =
    1 - q^2 \, x^3 - q^3 \, x^4 \, H^{(1)}(q \, x) ,
    \\[2mm]
    H^{(1)}(x)
    & =
    1 - q \, x^3 - q^6 \, x^8 \, H^{(0)}(q \, x) .
  \end{aligned}
\end{equation}

\begin{prop}
  \label{prop:H_function}
  Let the functions $H^{(a)}(x)$ be defined by eqs.~\eqref{define_H}.
  Then  for $a=0,1$  we have
  \begin{multline}
    \label{H_and_etc}
    H^{(1-a)}(x)
    =
    \sum_{n=0}^\infty (x)_{n+1} \, x^{2 n}
    \\
    \times
    \left(
      \sum_{k=0}^{\lfloor (n-a)/2 \rfloor} x^{2 k-a} q^{2 k(k+a)} \,
      \begin{bmatrix}
        n \\
        2 \, k+a
      \end{bmatrix}
      +
      \sum_{k=0}^{\lfloor (n+1-a)/2 \rfloor} x^{2 k+1-a} q^{2 k(k+a)} \,
      \begin{bmatrix}
        n+1 \\
        2 \, k+a
      \end{bmatrix}
    \right) .
  \end{multline}
\end{prop}

\begin{proof}
  We find that the periodic function~\eqref{define_chi} is related
  to the Dirichlet character
  \begin{equation*}
    -\chi_{24}^{(0)}(n) + \chi_{24}^{(1)}(n) =\left( \frac{12}{n} \right)
    \equiv \chi_{12}(n)  ,
  \end{equation*}
  where we have used the Legendre symbol.
  We further introduce a $q$-series $H(x; q)$ as
  \begin{align}
    -q^{\frac{1}{2}} \, x^2 \, H^{(0)}(x) + H^{(1)}(x)
    & =
    \sum_{n=0}^\infty \chi_{12}(n) \, q^{\frac{n^2-1}{48}} \,
    x^{\frac{n-1}{2}} 
    \nonumber     \\
    & \equiv
    H(x ; q^{\frac{1}{2}}).
    \label{define_Zagier}
  \end{align}
  As the functions  $H^{(a)}(x)$  have integral powers of $q$,
  the $q$-hypergeometric expression  may be derived if we know 
  that  of  $H(x; q^{\frac{1}{2}})$.
  Fortunately
  it is known from Refs.~\citen{Andre76,DZagie01a}
  that
  \begin{equation}
    H(x ; q^{\frac{1}{2}})
    =
    \sum_{n=0}^\infty (x ; q^{\frac{1}{2}})_{n+1} \, x^n .
  \end{equation}
  To divide this expression into a sum of
  $H^{(a)}(x)$, we compute as follows;
  \begin{align*}
    & \sum_{n=0}^\infty (x ; q^{\frac{1}{2}})_{n+1} \, x^n 
    \\
    & =
    \sum_{n=0}^\infty (x)_{n+1}  (x \, q^{\frac{1}{2}} )_n \, x^{2 n}    
    +
    \sum_{n=0}^\infty (x)_{n+1}  (x \, q^{\frac{1}{2}} )_{n+1} \, x^{2 n+1}
    \\
    & =
    \sum_{n=0}^\infty (x)_{n+1}  \, x^{2 n} \,
    \left(
      \sum_{j=0}^n 
      \begin{bmatrix}
        n \\
        j
      \end{bmatrix}
      \, (- 1 )^j \, x^j \, q^{j^2/2}
      +
      \sum_{j=0}^{n+1} 
      \begin{bmatrix}
        n+1 \\
        j
      \end{bmatrix}
      \, (- 1 )^j \, x^{j+1} \, q^{j^2/2}
    \right)
    \\
    & =
    \sum_{n=0}^\infty (x)_{n+1} \, x^{2 n}
    \sum_{k=0}^{\lfloor (n+1)/2 \rfloor}
    \left(
      \begin{bmatrix}
        n \\
        2 \, k
      \end{bmatrix}
      +
      x \,
      \begin{bmatrix}
        n +1 \\
        2 \, k
      \end{bmatrix}
    \right) \, x^{2 k} \, q^{2 k^2}
    \\
    & \qquad
    -q^{\frac{1}{2}} \, x^2 \,
    \sum_{n=0}^\infty (x)_{n+1} \, x^{2 n-1}
    \sum_{k=0}^{\lfloor n/2 \rfloor}
    \left(
      \begin{bmatrix}
        n \\
        2 \, k +1
      \end{bmatrix}
      +
      x \,
      \begin{bmatrix}
        n +1 \\
        2 \, k +1
      \end{bmatrix}
    \right) \, x^{2 k} \, q^{2 k (k+1)} .
  \end{align*}
  In the second equality, we have used
  eq.~\eqref{q_product_and_Gauss}.
  We have separated a sum into even and odd parts in both
  the first and the
  last equalities.
  This proves a statement of the proposition.
\end{proof}

Using the $q$-binomial formula~\eqref{q_binomial}, we can rewrite
eq.~\eqref{H_and_etc} into
\begin{multline}
  \label{H_lemma}
  H^{(1-a)}(x)
  =
  (q \, x)_\infty
  \sum_{n=0}^\infty
  \frac{q^{2 n (n + a)}}{(x^2 \, q)_{2 n + a}} \, x^{6 n + a - 1}
  \\
  + (1-x) \sum_{n=0}^\infty
  \left( (q \, x)_n - (q \, x)_\infty \right) \, x^{2 n}
  \sum_{k\geq 0}
  x^{2 k- a} \, q^{2 k (k + a)} \,
  \left(
    \begin{bmatrix}
      n \\
      2 \, k+a
    \end{bmatrix}
    + x \,
    \begin{bmatrix}
      n+1 \\
      2 \, k +a
    \end{bmatrix}
  \right) .
\end{multline}
In a limit $x\to 1$,
functions $H^{(a)}(x)$
defined by
eqs.~\eqref{define_H} can be written in an infinite product form with
a help of the Watson quintuple identity~\eqref{quintuple}.
Combining with a result~\eqref{H_lemma} in a limit $x\to 1$,
we recover  the following identities.

\begin{coro}[Slater's identities, (83) and~(86)~\cite{LJSlat52}]
  \begin{subequations}
    \begin{align}
      (q)_\infty \sum_{n=0}^\infty \frac{q^{2 n (n+1)}}{(q)_{2n+1}}
      & =
      (q^3 , q^5, q^8 ; q^8)_\infty \cdot
      (q^2 , q^{14} ; q^{16})_{\infty} ,
      \\[2mm]
      (q)_\infty \sum_{n=0}^\infty \frac{q^{2 n^2 }}{(q)_{2n}}
      & =
      (q , q^7, q^8 ; q^8)_\infty \cdot
      (q^6 , q^{10} ; q^{16})_{\infty}  .
    \end{align}
  \end{subequations}
\end{coro}

We obtain the following formulae
which come from the next order of $x-1$ of eq.~\eqref{H_lemma}.

\begin{prop}
  We have the following  $q$-series identities;
  \begin{subequations}
    \label{q_series_1}
    \begin{multline}
      \frac{1}{2} \sum_{n=0}^\infty n \, \chi_{24}^{(0)}(n) \,
      q^{\frac{n^2-25}{48}}
      \\
      =
      (q^3 ,q^5, q^8; q^8)_\infty \,
      (q^2 , q^{14} ; q^{16})_\infty \,
      \left(
        \sum_{k=1}^\infty \frac{- q^k}{1-q^k}
      \right)
      + (q)_\infty \sum_{n=0}^\infty \frac{q^{2 n (n+1)}}{(q)_{2 n+1}} \,
      \left(
        6 \, n+ \sum_{k=1}^{2 n+1} \frac{2 \, q^k}{1-q^k}
      \right)
      \\
      -
      \sum_{n=0}^\infty
      \left( (q)_n - (q)_\infty \right) \,
      \left( T_n(q) + T_{n+1}(q) \right) ,
    \end{multline}
    \begin{multline}
      \frac{1}{2} \sum_{n=0}^\infty n \, \chi_{24}^{(1)}(n) \,
      q^{\frac{n^2-1}{48}}
      \\
      =
      (q,q^7, q^8; q^8)_\infty \,
      (q^6 , q^{10} ; q^{16})_\infty \,
      \left(
        -1 - \sum_{k=1}^\infty \frac{q^k}{1-q^k}
      \right)
      + (q)_\infty \sum_{n=0}^\infty \frac{q^{2 n^2}}{(q)_{2 n}} \,
      \left(
        6 \, n+ \sum_{k=1}^{2 n} \frac{2  \, q^k}{1-q^k}
      \right)
      \\
      -
      \sum_{n=0}^\infty
      \left( (q)_n - (q)_\infty \right) \,
      \left( S_n(q) + S_{n+1}(q) \right) .
    \end{multline}
  \end{subequations}
\end{prop}

\begin{proof}
  We differentiate eq.~\eqref{H_lemma} with respect to $x$, and then
  substitute $x\to 1$.
\end{proof}

\begin{proof}[Proof of Theorem~\ref{theorem:Slater}]
  We substitute $q=\mathrm{e}^{-z}$ for  eqs.~\eqref{q_series_1}.
  As  terms which include
  infinite product terms such as $(q)_\infty$ vanish  in a
  limit $z\searrow 0$,
  we  get  formal $q$-series identities;
  \begin{subequations}
    \begin{align}
      X^{(0)}( q )
      & =
      - \frac{1}{2} \sum_{n=0}^{\infty} n \, \chi_{24}^{(0)}(n) \,
      q^{\frac{n^2-25}{48}} ,
      \\[2mm]
      X^{(1)}( q )
      & =
      - \frac{1}{2} \sum_{n=0}^{\infty} n \, \chi_{24}^{(1)}(n) \,
      q^{\frac{n^2-1}{48}}  .
    \end{align}
  \end{subequations}
  Applying Lemma~\ref{lemma:L_function}, we obtain the statement of Theorem.
\end{proof}


We see that
\begin{equation*}
  \chi_{24}^{(0)}(n) + \chi_{24}^{(1)}(n) = \left( \frac{24}{n} \right) \equiv \chi_{24}(n),
\end{equation*}
where  we have used the Legendre symbol.
Using  Zagier's result, we can see that,
when
the $q$-series $X(q)$ is defined by
  \begin{equation}
    X(q)
    =
    \sum_{n=0}^\infty
    (-q^\frac{1}{2} ; -q^{\frac{1}{2}})_n  ,
  \end{equation}
  we have an asymptotic expansion
  \begin{equation}
    X(\mathrm{e}^{-z})
    =
    \mathrm{e}^{z/48}
    \sum_{n=0}^\infty
    \frac{t_n}{n!} \,
    \left(
      \frac{z}{48}
    \right)^n ,
  \end{equation}
  where
  \begin{align*}
    t_n 
    & =
    \frac{1}{2} (-1)^{n+1} \,  L(-2 \, n -1, \chi_{24})
    \\
    &    = t_n^{(0)} + t_n^{(1)} .
  \end{align*}

\section{Modularity and Knot Invariant}
\label{sec:Knot}

The $q$-series which we have studied in preceding sections
is related to the modular form as follows.
We define
\begin{equation}
  \boldsymbol{\Phi}(\tau)
  \equiv
  \begin{pmatrix}
    \Phi^{(0)}(\tau) \\[2mm]
    \Phi^{(1)}(\tau) \\[2mm]
    \Phi^{(2)}(\tau) 
  \end{pmatrix}
  =
  \begin{pmatrix}
    q^{\frac{25}{48}} \, H^{(0)}(x=1) \\[2mm]
    q^{\frac{1}{48}} \, H^{(1)}(x=1) \\[2mm]
    \eta(2 \, \tau)
  \end{pmatrix} .
\end{equation}
Here
$q=\exp(2 \, \pi \, \mathrm{i} \, \tau)$
with $\tau$ in the upper half plane $\mathbb{H}$, and we have used the
Dedekind $\eta$-function,
\begin{equation}
  \eta(\tau) =
  q^{\frac{1}{24}} \, (q)_\infty
  =
  \sum_{n =0}^\infty \chi_{12}(n) \, q^{\frac{n^2}{24}} .
\end{equation}
We can find  by use of the Poisson summation formula
that
the vector $\boldsymbol{\Phi}(\tau)$
is  modular  with weight $1/2$;
the modular
$S$- and $T$-transformations are respectively written as
\begin{gather}
  \boldsymbol{\Phi}(\tau)
  =
  \sqrt{ \frac{ \  \mathrm{i} \  }{ \tau} } \,
  \begin{pmatrix}
    \frac{1}{2} & \frac{1}{2} & -\frac{1}{\sqrt{2}} \\[2mm]
    \frac{1}{2} & \frac{1}{2} & \frac{1}{\sqrt{2}} \\[2mm]
    - \frac{1}{\sqrt{2}} & \frac{1}{\sqrt{2}} & 0
  \end{pmatrix}
  \cdot
  \boldsymbol{\Phi}(-1/\tau)
  \equiv
  \sqrt{ \frac{ \  \mathrm{i} \  }{ \tau} } \,
  \mathbf{M} \cdot   \boldsymbol{\Phi}(-1/\tau) ,
  \label{Phi_under_S}
  \\[2mm]
  \boldsymbol{\Phi}(\tau+1)
  =
  \begin{pmatrix}
    \mathrm{e}^{\frac{25}{24} \pi \mathrm{i}} & & \\
    &\mathrm{e}^{\frac{1}{24} \pi \mathrm{i}} & \\
    & & \mathrm{e}^{\frac{1}{6} \pi \mathrm{i}}
  \end{pmatrix}
  \,
  \boldsymbol{\Phi}(\tau)  .
\end{gather}
These modular forms represent the character of the minimal model
$\mathcal{M}(3,4)$, \emph{i.e.},
the Ising model (see, \emph{e.g.},
Refs.~\citen{ItzyZube86,Rocha84a}.).

We consider an asymptotic behavior of $X^{(a)}(q)$ when $q$ is near at
the $N$-th primitive root of unity.
Hereafter we use
\begin{equation*}
  \omega= \mathrm{e}^{2 \pi \mathrm{i}/N} .
\end{equation*}
We define the Eichler integral of $\boldsymbol{\Phi}(\alpha)$ for
$\alpha \in \mathbb{Q}$ by
\begin{equation}
  \widetilde{\boldsymbol{\Phi}}(\alpha)
  \equiv
  \begin{pmatrix}
    \widetilde{\Phi}^{(0)}(\alpha ) \\
    \widetilde{\Phi}^{(1)}(\alpha ) \\
    \widetilde{\eta}(2 \, \alpha )
  \end{pmatrix}
  =
  \begin{pmatrix}
    \mathrm{e}^{\frac{25}{24} \pi \mathrm{i} \alpha } \,
    X^{(0)}(\mathrm{e}^{2 \pi \mathrm{i} \alpha })
    \\
    \mathrm{e}^{\frac{1}{24} \pi \mathrm{i} \alpha } \,
    X^{(1)}(\mathrm{e}^{2 \pi \mathrm{i} \alpha })
    \\
    \mathrm{e}^{\frac{1}{6} \pi \mathrm{i} \alpha} \,
    X^{(2)}(\mathrm{e}^{2 \pi \mathrm{i} \alpha })
  \end{pmatrix} ,
\end{equation}
where $X^{(0)}(q)$ and $X^{(1)}(q)$ are defined in
eqs.~\eqref{define_X}, and
\begin{equation}
  X^{(2)}(q)
  =
  2   \sum_{n=0}^\infty (q^2 ; q^2)_n .
\end{equation}
One sees that
$\widetilde{\boldsymbol{\Phi}}(\alpha)$ converges to finite value
for $\alpha \in \mathbb{Q}$
as an infinite sum terminates at finite order
due to $(q)_n$.

As was proved in Ref.~\citen{KHikami03c} (see also
Refs.~\citen{LawrZagi99a,DZagie01a}), we have an asymptotic behavior
of the Eichler integral of the modular form of weight $1/2$.
\begin{theorem}[\cite{KHikami03c}]
  For $N\in \mathbb{Z}_{>0}$, we have an
  asymptotic expansion in $N\to \infty$ as
  \begin{equation}
    \label{modular_Phi_tilde}
    \widetilde{\boldsymbol{\Phi}}(1/N)
    +
    \left( - \mathrm{i} \, N \right)^{\frac{3}{2}} \, \mathbf{M} \cdot
    \widetilde{\boldsymbol{\Phi}}(-N)
    \simeq
    \sum_{n=0}^\infty \frac{\mathbf{t}_n}{n!} \,
    \left(
      \frac{\pi}{24 \, \mathrm{i} \, N}
    \right)^n ,
  \end{equation}
  where $\mathbf{M}$ is the $3\times 3$ matrix
  defined in eq.~\eqref{Phi_under_S}.
  We mean that  $t$-series is
  \begin{equation*}
    \mathbf{t}_n
    =
    \begin{pmatrix}
      t_n^{(0)} \\
      t_n^{(1)} \\
      t_n^{(2)}
    \end{pmatrix} ,
  \end{equation*}
  where $t_n^{(0)}$ and $t_n^{(1)}$ are given in
  eq.~\eqref{define_t_12}, and
  \begin{align}
    \label{define_t_etc}
    t_n^{(2)}
    & = -(-4)^n \, L(-2 \, n -1 , \chi_{12}) 
    \nonumber
    \\
    &=
    \frac{1}{2} \, (-1)^n \, 
    \frac{24^{2 n+1}}{2 \, n+2} \,
    \sum_{k=1}^{12} \chi_{12}(k) \, B_{2 n+2}(k/12) .
  \end{align}
\end{theorem}

It is  noted that the Eichler integrals
$\widetilde{\boldsymbol{\Phi}}(N)$
at $N \in \mathbb{Z}$  in the left hand side of
eq.~\eqref{modular_Phi_tilde}
are computed as 
\begin{equation}
  \label{Eichler_at_integer}
  \widetilde{\boldsymbol{\Phi}}(N)
  =
  \begin{pmatrix}
    \mathrm{e}^{\frac{25}{24} \pi \mathrm{i} N} \\[2mm]
    2 \, \mathrm{e}^{\frac{1}{24} \pi \mathrm{i} N} \\[2mm]
    2 \, \mathrm{e}^{\frac{1}{6} \pi \mathrm{i} N} 
  \end{pmatrix}.
\end{equation}


\begin{figure}[htbp]
  \centering

  \includegraphics{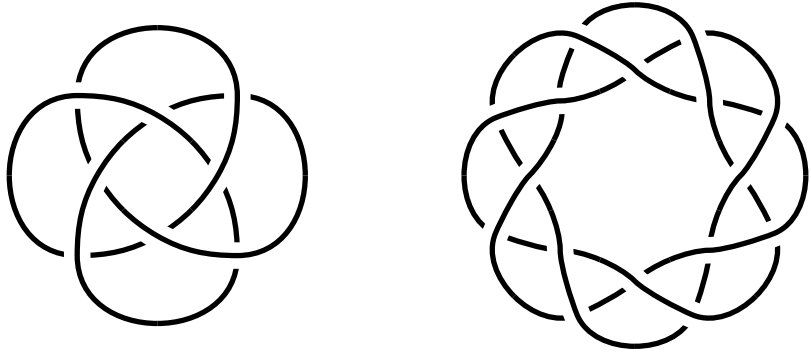}
  \caption{Torus knots, $\mathcal{T}_{3,4}$ and $\mathcal{T}_{3,8}$.}
  \label{fig:torus34proj}
\end{figure}

\begin{prop}[\cite{KHikami03c}]
  \label{prop:Kashaev_34}
  Kashaev's invariant $\langle \mathcal{K} \rangle_N$
  for torus knot
  $\mathcal{K}=\mathcal{T}_{3,4}$
  (see Fig.~\ref{fig:torus34proj})
  is proportional to
  $X^{(0)}(\omega)$;
  \begin{equation}
    \label{Ising_invariant}
    \omega^{3} \, X^{(0)}(\omega)
    =
    \langle \mathcal{T}_{3,4} \rangle_N  .
  \end{equation}
\end{prop}

See also Ref.~\citen{KHikami04a} for a connection with the colored Jones polynomial with generic $q$,
the Alexander polynomial,
A-polynomial,
for the torus knot.

The quantum invariant such as the colored Jones polynomial
was identified with the Chern--Simons path integral~\cite{EWitt89a}.
The asymptotic behavior thereof is known to be related to the classical
topological invariants such as the Reidemeister torsion and the Chern--Simons
invariant.
Looking at the nearly modular property~\eqref{modular_Phi_tilde}
with eq.~\eqref{Ising_invariant},
the limiting value of the Eichler
integral~\eqref{Eichler_at_integer} at $\tau \to N\in \mathbb{Z}$
can be identified with
the Chern--Simons invariant of the torus knot
$\mathcal{T}_{3,4}$~\cite{KHikami03c}.

\section{\mathversion{bold}  Applications}
\label{sec:application}

In our previous paper~\cite{KHikami03c}, we have demonstrated that
Kashaev's invariant for the torus knot
$\mathcal{T}_{s,t}$ where
$s$ and $t$ are relatively prime positive integers
(see \emph{e.g.} Fig.~\ref{fig:torus34proj})
is regarded as the Eichler integral of the Virasoro character
of the minimal model $\mathcal{M}(s,t)$.
Prop.~\ref{prop:Kashaev_34} is for  a case of $(s,t)=(3,4)$.
The Virasoro character $\ch_{n,m}^{s,t}(\tau)$  of $\mathcal{M}(s,t)$
for an irreducible highest weight with conformal weight
$\Delta_{n,m}^{s,t}
=
\frac{( n t - m s)^2-(s-t)^2}{4 s t}
$
for $1 \leq n \leq s-1$ and $1 \leq m \leq t-1$,
is known to be~\cite{Rocha84a}
\begin{equation}
  \ch_{n,m}^{s,t}(\tau)
  =
  \frac{\Phi_{s,t}^{(n,m)}(\tau)}{\eta(\tau)} .
\end{equation}
Here we have
\begin{equation}
  \Phi_{s,t}^{(n,m)}(\tau)
  = \sum_{k=0}^\infty
  \chi_{2 s t}^{(n,m)}(k) \,
  q^{\frac{k^2}{4 s t}} ,
\end{equation}
with a periodic function
\begin{equation*}
  \begin{array}{c|ccccc}
    k \mod 2 \, s \, t &
    n \, t - m \, s & n \, t + m \, s &
    2 \, s \, t - ( n \, t+ m \,s) &
    2 \, s \, t - (n \, t - m \, s) &
    \text{others}
    \\
    \hline
    \chi_{2 s t}^{(n,m)}(k) &
    1 & -1 & -1 & 1 & 0
  \end{array}
\end{equation*}
The function $\Phi_{s,t}^{(n,m)}(\tau)$ is modular covariant with
weight $1/2$~\cite{ItzyZube86,CappItzyZube87b}, and
spans a $(s-1)\, (t-1)/2$-dimensional space due to a symmetry
$\ch_{n,m}^{s,t}(\tau)
=\ch_{s-n,t-m}^{s,t}(\tau)$.
Then Kashaev's invariant $\langle \mathcal{K} \rangle_N$ for a torus
knot $\mathcal{K}=\mathcal{T}_{s,t}$ is written in terms of the
Eichler integral $\widetilde{\Phi}_{s,t}^{(1,t-1)}(1/N)$,
which
is defined by the half-differential of the modular form as
\begin{equation}
  \widetilde{\Phi}_{s,t}^{(n,m)}(\tau)
  =
  -\frac{1}{2} \sum_{k=0}^\infty
  k \, \chi_{2 s t}^{(n,m)}(k) \,
  q^{\frac{k^2}{4 s t}} ,
\end{equation}
where $\tau\in \mathbb{H}$.
The limiting value at $\tau\to 1/N\in\mathbb{Q}$ is computed as
\begin{equation}
  \widetilde{\Phi}_{s,t}^{(n,m)}(1/N)
  =
  \frac{s \, t \, N}{2}
  \sum_{k=1}^{2 s t N}
  \chi_{2 s t}^{(n,m)}(k) \,
  \mathrm{e}^{\frac{k^2}{2 s t N} \pi \mathrm{i}} \,
  B_2
  \left(
    \frac{k}{2 \,s \, t \, N}
  \right).
\end{equation}

We consider the $q$-series identities associated with (the Eichler integral
of) the minimal model $\mathcal{M}(3,k)$.
As we have discussed in preceding sections, the character
of the minimal model $\mathcal{M}(3,4)$ is related to
Slater's identities, and we can expect
that such identities could be constructed.
In fact
there are multi-variable generalizations of
Slater's identities due to
G.~Andrews\cite{Andre84a}.
These $q$-series identities may be interpreted as fermionic formulae for the  characters 
of the Virasoro algebra.
The $q$-series identities which come from the computation of the quantum invariants of the corresponding torus knots
$\mathcal{T}_{3,k}$ are rather messy.

To study  $q$-series identities associated to the character of the
minimal model
$\mathcal{M}(s,t)$, it will be convenient to define 
a function~\cite{KHikami03c,KHikami04a}
\begin{equation}
  \label{function_of_x}
  H_{s , t}^{(n,m)}(x)
  \equiv
  H_{s,  t}^{(n,m)}(x ; q)
  =
  \sum_{k=0}^\infty
  \chi_{2 s t}^{(n,m)}(k) \,
  q^{
    \frac{k^2 - (n  t - m s)^2}{4 s t}
  } \,
  x^{
    \frac{k -
      \left|  n t - m s \right|
    }{2}
  } .
\end{equation}
See that
\begin{equation*}
  \Phi_{s,t}^{(n,m)}(\tau)
  =
  q^{\frac{( n t - m s)^2}{4 s t}} \,
  H_{s, t}^{(n,m)}(1 ) ,
\end{equation*}
and that $H^{(a)}(x)$ for $a=0,1$
defined in eq.~\eqref{define_H} is
nothing but
$H_{s,  t}^{(n,m)}(x)$ with
$(s,t)=(3,4)$ and
$(n,m)=(1,3), (1,1)$, respectively.
We  demonstrate  below how our function~\eqref{function_of_x} does
work
for
investigations of the $q$-hypergeometric type functions associated to the
minimal Virasoro characters.
It is noted
that linear  combinations  among  characters for minimal Virasoro model was
studied in Ref.~\citen{BytsFrin00a} in a different method.


\subsection{\mathversion{bold}
  $\mathcal{M}(3,2^p)$}
First we pay attention to a case of $(s,t)=(3, 2 \, k)$.
The periodic function  satisfies
\begin{equation}
  \chi_{12} (2 \, k +3) \,
  \chi_{12}(n)
  =
  \sum_{a=0}^{k-1} (-1)^{a-1} \,
  \chi_{12 k}^{(1,2 a + 1)}(n) .
\end{equation}
Using this identity, we have the  following proposition;
\begin{prop}
  Let the function $H(x; q)$ be defined by eq.~\eqref{define_Zagier}.
  We have
\begin{equation}
  \chi_{12}(2 \, k + 3) \,
  H(x; q^{\frac{1}{k}})
  =
  \sum_{a=0}^{k-1}
  (-1)^{a-1} \,
  q^{\ell_k(a)}
  x^{\frac{|2 k - 6 a - 3 | - 1}{2}}
  \,
  H_{3, 2 k}^{(1 , 2 a +1)}(x; q) ,
\end{equation}
where
\begin{equation*}
  \ell_k(a) = \frac{(k-3 \, a -1) (k - 3 \, a -2)}{6 \, k} .  
\end{equation*}
\end{prop}

When $k=2^p$, we see that $\ell_k(a) - \ell_k(b) \not\in \mathbb{Z}$
for all $a$ and $b$ satisfying
$0 < a \neq b \leq 2 \, k-1$.
Thus we can extract 
$H_{3 , 2^{p+1}}^{(1 , 2 a +1)}(x; q) $
from above lemma
for a case of $\mathcal{M}(3,2^{p+1})$
following
a method of Prop.~\ref{prop:H_function}.


We take an example $\mathcal{M}(3,8)$.
We have
\begin{subequations}
  \begin{align}
    H(x ; q^{\frac{1}{4}})
    & =
    H_{3,8}^{(1,3)}(x) - q^{\frac{1}{4}} \, x^2 \, H_{3,8}^{(1,1)}(x)
    - q^{\frac{1}{2}}  x^3 \, H_{3,8}^{(1,5)}(x)
    + q^{\frac{7}{4}} \, x^6 \, H_{3,8}^{(1,7)}(x) ,
    \label{48_and_H}
    \\[2mm]
    H_{3,8}^{(1,2)}(x )
    & =
    H^{(1)}(x^2 ; q^2 ) ,
    \\[2mm]
    H_{3,8}^{(1,6)}(x)
    & =
    H^{(0)}(x^2 ; q^2) ,
    \\[2mm]
    H_{3,8}^{(1,4)}(x)
    & =
    H(x^4 ; q^4) .
  \end{align}
\end{subequations}
{}From a result in Section~\ref{sec:proof}
we already have a $q$-hypergeometric 
expression of $H_{3,8}^{(1,a)}(x)$ for even $a$ cases.
For odd $a$ cases, we can derive   $q$-hypergeometric functions from
eq.~\eqref{48_and_H}  by the  same method with previous section.
We have
\begin{align*}
  H_{3,8}^{(1,1)}(x)
  & =
  - q^{-\frac{1}{4}} \, x^{-2} \, I^{(1)}(x) ,
  \\[2mm]
  H_{3,8}^{(1,3)}(x)
  & = I^{(0)}(x) ,
  \\[2mm]
  H_{3,8}^{(1,5)}(x)
  & =
  - q^{- \frac{1}{2}} \, x^{-3} \, I^{(2)}(x) ,
  \\[2mm]
  H_{3,8}^{(1,7)}(x)
  & =
  q^{- \frac{7}{4}} \, x^{-6} \, I^{(3)}(x) .
\end{align*}
where
\begin{multline*}
  I^{(a)}(x)
  =
  \sum_{k=0}^\infty (x)_{k+1} \, x^{4 k} \, \sum_{1 \geq \epsilon_1 \geq \epsilon_2 \geq \epsilon_3 \geq 0}
  x^{\epsilon_1+\epsilon_2+\epsilon_3}
  \\
  \times
  \sum_{
    \substack{
      \ell_1, \ell_2,  \ell_3 \geq 0 \\
      \ell_1 + 2 \ell_2 + 3 \ell_3 = a \mod 4
    }}
  (-x)^{\ell_1 + \ell_2 + \ell_3} \, q^{\frac{\ell_1 + 2 \ell_2 + 3 \ell_3}{4}} \,
  \left(
    \prod_{j=1}^3
    \begin{bmatrix}
      k+ \epsilon_j \\
      \ell_j
    \end{bmatrix}
    \,
    q^{\frac{1}{2} \ell_j ( \ell_j -1)} 
  \right) .
\end{multline*}
\subsection{\mathversion{bold}
  $\mathcal{M}(3,5)$}

We list  some formulae
concerning $H_{s,  t}^{(n,m)}(x)$   for $(s,t)=(3,5)$ defined in
eq.~\eqref{function_of_x}.
In this case we have four independent functions $(n,m)=(1,1),
(1,2),(1,3)$ and $(1,4)$, and
they satisfy the following difference equations;
\begin{gather*}
  H_{3,5}^{(1,4)}(x)
  = 1 - q^2 \, x^3 - q^4 \, x^5 \cdot H_{3,5}^{(1,1)}(q \, x) ,
  \\[2mm]
  H_{3,5}^{(1,3)}(x)
  = 1 - q^3 \, x^5 - q^4 \, x^6 \cdot H_{3,5}^{(1,2)}(q \, x) ,
  \\[2mm]
  H_{3,5}^{(1,2)}(x)
  = 1 - q^2 \, x^5 - q^6 \, x^9 \cdot H_{3,5}^{(1,3)}(q \, x) ,
  \\[2mm]
  H_{3,5}^{(1,1)}(x)
  = 1 - q \, x^3 - q^8 \, x^{10} \cdot H_{3,5}^{(1,4)}(q \, x) ,
\end{gather*}
We note that
\begin{gather*}
  -q^{\frac{3}{4}} \, x^{\frac{5}{2}} \, H_{3,5}^{(1,4)}(x)
  + H_{3,5}^{(1,1)}(x)
  =
  H_{3,5}^{(1,3)}(x^{\frac{1}{2}} ; q^{\frac{1}{4}}) ,
  \\[2mm]
  -q^{\frac{1}{4}} \, x^{\frac{3}{2}} \, H_{3,5}^{(1,3)}(x)
  + H_{3,5}^{(1,2)}(x)
  =
  H_{3,5}^{(1,1)}(x^{\frac{1}{2}} ; q^{\frac{1}{4}}) ,
\end{gather*}
and that
the Watson quintuple identity~\eqref{quintuple}
proves
\begin{align*}
  H_{3,5}^{(1,4)}(1)
  & =
  (q^2, q^{18}; q^{20})_\infty \,
  (q^4, q^6 ,  q^{10}; q^{10})_\infty ,
  \\[2mm]
  H_{3,5}^{(1,3)}(1)
  & =
  (q^4, q^{16}; q^{20})_\infty \,
  (q^3, q^7 ,  q^{10}; q^{10})_\infty ,
  \\[2mm]
  H_{3,5}^{(1,2)}(1)
  & =
  (q^6, q^{14}; q^{20})_\infty \,
  (q^2, q^8 ,  q^{10}; q^{10})_\infty ,
  \\[2mm]
  H_{3,5}^{(1,1)}(1)
  & =
  (q^8, q^{12}; q^{20})_\infty \,
  (q, q^9 ,  q^{10}; q^{10})_\infty .
\end{align*}

Though we do not obtain a $q$-series identity which appeared in 
Slater's $130$ identities,
we obtain
the following formulae.

\begin{prop}
\begin{align}
  H_{3,5}^{(1,4)}(x)
  & =
  \sum_{n=0}^\infty (-q^4 \, x^5; q^{10})_n \,
  \left( - q^4 \, x^5 \right)^n
  -
  q^2 \, x^3
  \sum_{n=0}^\infty (-q^6 \, x^5; q^{10})_n \,
  \left( - q^6 \, x^5 \right)^n
  \label{identity_H^0_4}
  \\
  & =
  \sum_{n=0}^\infty (-q^2 \, x^3; q^6)_n \,
  \left( -q^2 \, x^3 \right)^{n+2c}
  \sum_{c=0}^n
  q^{6 c^2} \,
  \begin{bmatrix}
    n \\
    c
  \end{bmatrix}_{q^6}
  \nonumber \\
  & \qquad \qquad
  -
  q^4 \, x^5 \,
  \sum_{n=0}^\infty (-q^4 \, x^3; q^6)_n \,
  \left( -q^4 \, x^3 \right)^{n+2c}
  \sum_{c=0}^n
  q^{6 c^2} \,
  \begin{bmatrix}
    n \\
    c
  \end{bmatrix}_{q^6}
  \label{identity_H^0_1}
  \\
  & =
  \sum_{n,c=0}^\infty
  (-1)^n q^{\frac{5}{2} n(n+1) + 5 c(c+1)-n-c} \,
  x^{5(n+c)} \,
  \left(
    1 - q^{2(n+c+1)} \, x^3
  \right) \,
  \begin{bmatrix}
    n \\
    c
  \end{bmatrix}_{q^5}
  \label{identity_H^0}
\end{align}


\begin{align}
  H_{3,5}^{(1,3)}(x)
  & =
  - q^3 \, x^5
  \sum_{n,c =0}^\infty
  (-q \, x^3 ; q^6 )_{n+1} \,
  \left( - q \, x^3  \right)^{n+2c} \,
  q^{6 c (c+1)} \,
  \begin{bmatrix}
    n \\
    c
  \end{bmatrix}_{q^6}
  \nonumber \\
  & \qquad
  +
  \sum_{n,c =0}^\infty
  (-q^{-1} \, x^3 ; q^6 )_{n+1} \,
  \left( - q^{-1} \, x^3  \right)^{n+2c} \,
  q^{6 c (c+1)} \,
  \begin{bmatrix}
    n \\
    c
  \end{bmatrix}_{q^6}
  \label{identity_H^2-1}
  \\
  & =
  \sum_{n,c=0}^\infty
  (-1)^n q^{\frac{5}{2} n(n+1) + 5 c(c+1)-2(n+c)} \,
  x^{5(n+c)} \,
  \left(
    1 - q^{4(n+c+1)} \, x^6
  \right) \,
  \begin{bmatrix}
    n \\
    c
  \end{bmatrix}_{q^5}
  \label{identity_H^2-2}
\end{align}

\begin{align}
  H_{3,5}^{(1,2)}(x)
  & =
  \sum_{n=0}^\infty (-q^2 \, x^5 ; q^{10})_n \,
  \left( - q^2 \, x^5 \right)^n
  -
  q^6 \, x^9
  \sum_{n=0}^\infty (-q^8 \, x^5 ; q^{10})_n \,
  \left( - q^8 \, x^5 \right)^n
  \label{identity_H^3_2}
  \\
  & =
  \sum_{n,c=0}^\infty
  (-1)^n q^{\frac{5}{2} n(n+1) + 5 c(c+1)-3(n+c)} \,
  x^{5(n+c)} \,
  \left(
    1 - q^{6(n+c+1)} \, x^9
  \right) \,
  \begin{bmatrix}
    n \\
    c
  \end{bmatrix}_{q^5}
  \label{identity_H^3}
\end{align}

\begin{align}
  H_{3,5}^{(1,1)}(x)
  & =
  1-x^{-2} +
  x^{-2}
  \sum_{n,c=0}^\infty
  (-1)^n q^{\frac{5}{2} n(n+1) + 5 c(c+1)-4(n+c)} \,
  x^{5(n+c)} \,
  \left(
    1 - q^{8(n+c+1)} \, x^{12}
  \right) \,
  \begin{bmatrix}
    n \\
    c
  \end{bmatrix}_{q^5}
  \\
  & =
  - q \, x^3 \,
  \sum_{n=0}^\infty
  (-q \, x^5 ; q^{10})_{n+1} \,
  \left( - q \, x^5 \right)^n
  +
  \sum_{n=0}^\infty
  (-q^{-1} \, x^5 ; q^{10})_{n+1} \,
  \left( - q^{-1} \, x^5 \right)^n
  \label{identity_H^1}
\end{align}

\end{prop}

\begin{proof}
  We recall results from Refs.~\citen{DZagie01a,KHikami02c};
  \begin{gather}
    \sum_{n=0}^\infty \chi_6(n) \, q^{\frac{n^2-1}{24}}
    \,x^{\frac{n-1}{2}}
    =
    \sum_{n=0}^\infty
    (-x)_{n+1} \,(-x)^n ,
    \label{minus_Zagier_identity}
    \\[2mm]
    \sum_{n=0}^\infty \chi_{10}^{(0)}(n) \, q^{\frac{n^2-9}{40}} \,
    x^{\frac{n-3}{2}}
    =
    \sum_{n=0}^\infty
    (-x)_{n+1} \, (-x)^n
    \sum_{c=0}^n
    q^{c(c+1)} \, x^{2 c} \,
    \begin{bmatrix}
      n \\
      c
    \end{bmatrix}  ,
    \label{minus_Hikami_identity}
    \\[2mm]
    \sum_{n=0}^\infty \chi_{10}^{(1)}(n) \, q^{\frac{n^2-1}{40}} \,
    x^{\frac{n-1}{2}}
    =
    \sum_{n=0}^\infty
    (-x)_{n+1} \, (-x)^n
    \sum_{c=0}^{n+1}
    q^{c^2} \, x^{2 c} \,
    \begin{bmatrix}
      n +1 \\
      c
    \end{bmatrix} .
    \label{minus_Hikami_identity_2}
  \end{gather}
  Here the periodic functions are meant to be
  \begin{gather*}
    \begin{array}{c|ccc}
      n \mod 6 & 1 & 5 & \text{others}
      \\
      \hline      \hline
      \chi_{6}(n) & 1 & -1 & 0
    \end{array}
    \\[2mm]
    \begin{array}{c|ccc}
      n \mod 10 & 3 & 7 & \text{others}
      \\
      \hline      \hline
      \chi_{10}^{(0)}(n) & 1 & -1 & 0
    \end{array}
    \\[2mm]
    \begin{array}{c|ccc}
      n \mod 10 & 1 & 9 & \text{others}
      \\
      \hline      \hline
      \chi_{10}^{(1)}(n) & 1 & -1 & 0
    \end{array}
  \end{gather*}

  Using eq.~\eqref{minus_Zagier_identity},
  $\chi_{30}^{(1,1)}(n)
  =
  -\chi_6(\frac{n-3}{5})+\chi_6(\frac{n+3}{5})$,
  and
  $\chi_{30}^{(1,2)}(n)
  =
  -\chi_6(\frac{n-6}{5})+\chi_6(\frac{n+6}{5})$,
  we obtain eqs.~\eqref{identity_H^1} and~\eqref{identity_H^3_2}.
  Eq.~\eqref{identity_H^0_4} also follows in the same manner.
  Eqs.~\eqref{identity_H^0_1}  and~\eqref{identity_H^2-1} follow from
  eqs.~\eqref{minus_Hikami_identity}, \eqref{minus_Hikami_identity_2},
  $\chi_{30}^{(1,3)}(n)
  =
  -\chi_{10}^{(0)}(\frac{n-5}{3})+
  \chi_{10}^{(0)}(\frac{n+5}{3})$, and
  $\chi_{30}^{(1,4)}(n)
  =
  -\chi_{10}^{(1)}(\frac{n-10}{3})+
  \chi_{10}^{(1)}(\frac{n+10}{3})$.

  To prove remaining identities, we use a formula~\cite{KHikami03a},
  \begin{equation}
    \label{link_identity}
    \sum_{n=0}^\infty
    \widetilde{\chi}_6(n) \, q^{\frac{n^2-4}{12}} \,x^{\frac{n-2}{2}}
    =
    \sum_{n=0}^\infty (-1)^n \, q^{\frac{1}{2} n(n+1)}
    \sum_{c=0}^n
    x^{n+c} \,q^{c(c+1)} \,
    \begin{bmatrix}
      n \\
      c
    \end{bmatrix} ,
  \end{equation}
  with
  \begin{equation*}
    \begin{array}{c|ccc}
      n \mod 6 & 2 & 4 & \text{others}
      \\
      \hline      \hline
      \widetilde{\chi}_{6}(n) & 1 & -1 & 0
    \end{array}
  \end{equation*}
  Combining with a fact that
  $\chi_{30}^{(1,5-a)}(n) = - \widetilde{\chi}_6(\frac{n-3 \, a}{5})
  +
  \widetilde{\chi}_6(\frac{n+3 \, a}{5})$
  for $a=1,2,3$,
  we can complete the proof.
\end{proof}

We should remark that 
an identity~\eqref{link_identity} gives
\begin{equation}
  H_{3,t}^{(1,m)}(x)
  =
  \begin{cases}
    I_t^{(m)} (x) ,
    & \text{for $t - 3 \, m <0$},
    \\[2mm]
    1-x^{3 m - t} + x^{3 m -t} \,
    I_t^{(m)}(x) ,
    &
    \text{for $t - 3 \, m>0$} ,
  \end{cases}
\end{equation}
where
\begin{equation*}
  I_t^{(m)}(x)
  =
  \sum_{n,c=0}^\infty (-1)^n \,
  q^{\frac{t}{2} n (n+1) + t c ( c+1) -(t-m) (n+c)} \,
  x^{t (n+c)} \,
  \left(
    1 -q^{2 (t-m) ( c + n+1)} \, x^{3 (t-m)}
  \right) \,
  \begin{bmatrix}
    n \\
    c
  \end{bmatrix}_{q^t} .
\end{equation*}
Here we have set
$t>0$ and $1\leq m <t$ such that
$(3,t)=1 $,
and $H_{3,t}^{(1,m)}(x=1)$ is a ($t-1$)-dimensional representation of the modular group.

\section*{Acknowledgment}
The authors would like to thank J.~Kaneko and H.~Murakami for useful
discussions.
Work of KH is supported in part by
Grant-in-Aid for Young Scientists
from the Ministry of Education, Culture, Sports, Science and
Technology of Japan.


\end{document}